\documentclass[a4paper,11pt]{article}
\usepackage{pos}
\renewcommand{\logo}{\vbox{\hfill{\tt BU-HEPP-24-05}}\vspace{-2cm}}
\renewcommand{\posurl}{}
\title{Phenomenology of matching exponentiated photonic radiation to a QED-corrected parton shower in KKMChh}
\ShortTitle{Parton shower matching in KKMChh}
\author*[a]{Scott A.\ Yost}
\author[b]{B.F.L.\ Ward}
\author[c]{Zbigniew W{\c a}s}
\affiliation[a]{Physics Department, The Citadel,\\
  171 Moultrie St., Charleston, SC 29409, U.S.A.}
\affiliation[b]{Physics Department, Baylor University,\\
1311 S 5th St., Waco, TX 76706, U.S.A.}
\affiliation[c]{Institute of Nuclear Physics Polish Academy of Science,\\
ul. Radzikowskiego 152 31-342 Krak{\'o}w, Poland}
\emailAdd{scott.yost@citadel.edu}
\emailAdd{bfl\_ward@baylor.edu}
\emailAdd{zbigniew.was@ifj.edu.pl}
\abstract{KKMChh is a precision Monte Carlo program for photonic and electroweak radiative corrections to hadron scattering, implementing the amplitude-level exponentiation originally developed for electron-positron scattering at the quark level, modeling initial and final state QED radiation as well as initial-final interference to all orders in a soft-photon approximation, adding hard photon corrections through second order next-to-leading logarithm. A previous ICHEP talk introduced a matching procedure NISR (negative initial-state radiation) to match the exponentiated photon radiation from the quarks to a QED-corrected parton shower. Here, we describe its effect on forward-backward asymmetry calculations of interest for a precise determination of the electroweak mixing angle. }
\FullConference{42nd International Conference on High Energy Physics (ICHEP2024)\\
18-24 July 2024\\
Prague, Czech Republic\\}
\begin{document}
\maketitle
KKMChh~\cite{KKMChh:2016,KKMChh:2019} is a precision Monte Carlo program for photonic and electroweak radiative corrections to hadron scattering, implementing amplitude-level exponentiation for initial-state (ISR) and final-state (FSR) photonic radiation as well as initial-final interference (IFI) to all orders in a soft-photon approximation and adding hard photon corrections through order $\alpha^2$ NLL.

The PDF sets may already contain some QED effects. This can come from QED ``contamination'' in the input data for PDF sets including only QCD evolution or from photon momenta in QED-corrected PDF sets. The NISR (negative ISR) mechanism avoids double-counting QED radiation already included in the PDF set.~\cite{ichep2022} NISR is applied at a scale $q_0$ beyond which ISR must be backed out. For a QED-corrected PDF set, the starting scale for NISR is taken to be the hard process scale. For a standard PDF set, $q_0$ may be the starting point of QCD evolution. 

Here, we focus on the effect of NISR on the calculation of the forward-backward asymmetry $A_{\rm FB}$ of muon pairs in proton collisions at 8 TeV CM energy, previously calculated without NISR.~\cite{kkmchh-afb,kkmchh-afb-arxiv} We compare three cases. Case (1) is the calculation without NISR using standard NNPDF-3.1 (NLO).~\cite{nnpdf1} Case (2) uses NNPDF-3.1 (NLO) with NISR applied at the starting scale of QCD evolution, $q_0 = 2$ GeV and case (3) uses NNPDF-3.1-LuxQED (NLO)~\cite{nnpdf2} with NISR at the hard process scale.  No hadronic shower was included and no cuts were applied to the muons. The results are based on MC samples with 8-10 billion weighted events.

Table 1 shows $A_{\rm FB}$ for these three cases in five ranges of the muon pair invariant mass $M_{\mu^-\mu^+}$, together with the difference between each of the NISR cases and case (1) without NISR. Plots (a) and (b) in Figure 1 show the same results in finer $M_{\mu^-\mu^+}$ bins, with case (1) in green, case (2) in blue, and case (3) in red. The differences $(2) - (1)$ and $(3) - (1)$ are shown in plot (b). Plots (c) and (d) show the same calculations binned in the muon pair rapidity $Y_{\mu^-\mu^+}$ instead.

\begin{table}[h]
\begin{center}
\caption{$A_{\rm FB}$ calculated in $M_{\mu^-\mu^+}$ bins and the differences with and without NISR. The MC error in the last digit is shown in parentheses.}
\label{tab1}
\renewcommand{\arraystretch}{0.8}
\begin{tabular}{|c|r|r|r|r|r|}
\hline
 &{\small\bf 89--93 GeV}&{\small\bf 60--120 GeV}&{\small\bf 60--81 GeV }&{\small\bf 81--101 GeV}&{\small\bf 101--150 GeV}\\
\hline
{\small\bf Case (1)}&
\small 0.03092(1)& \small 0.02071(1)& \small $-$0.10869(4)& \small 0.02886(1)& \small 0.2186(1)\\
{\small\bf Case (2)}&
\small 0.03128(4)& \small 0.02109(4)& \small $-$0.10853(9)& \small 0.02925(3)& \small 0.2192(1)\\
{\small\bf Case (3)}&
\small 0.03135(6)& \small 0.02115(4)& \small $-$0.1088(1)& \small 0.02934(5)& \small 0.2198(2) \\
\hline
{\small\bf Difference (2) $\mathbf -$ (1)}&
\small 0.00036(4)& \small 0.00038(4)& \small 0.0002(1)& \small 0.00038(3)& \small 0.0007(2)\\
{\small\bf Difference (3) $\mathbf -$ (1)}&
\small 0.00043(6)& \small 0.00044(5)& \small $-$0.0001(1)& \small 0.00047(5)& \small 0.0012(2)\\ \hline
\end{tabular}
\end{center}
\end{table}

In plots (a) and (b), the effect of NISR is small and mostly compatible within the MC errors for both ways of adding NISR.  A similar pattern of small NISR corrections, mostly compatible within MC errors, is seen in the table. The contribution of NISR to the rapidity plots is more significant and increases for large rapidity. Both cases of adding NISR are indistinguishable on the rapidity plots.

\begin{figure}[ht]
\begin{center}
    \includegraphics[width=\textwidth]{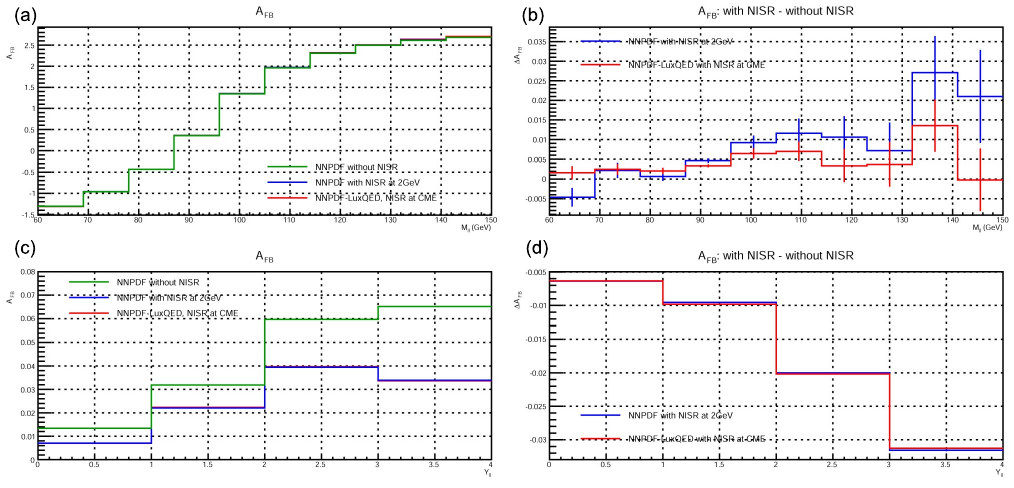}
    \caption{Plots (a) and (c) show $A_{\rm FB}$ as a function of the muon pair invariant mass $M_{\mu^-\mu^+}$ and rapidity $Y_{\mu^-\mu^+}$, for the three cases. Plots (b) and (d) show the corresponding differences $(2) - (1)$ and $(3) - (2)$. }\label{fig1}
\end{center}
\end{figure}

\acknowledgments

We acknowledge the essential contributions of Stanisław Jadach (1947 -- 2023), a creator of KKMC and leader in the field of precision radiative corrections.
S.A.\ Yost was supported in part by a grant from the Citadel Foundation.

\end{document}